# Sensitivity-Tunable Terahertz Liquid/Gas Biosensor Based on Surface Plasmon Resonance with Dirac Semimetal

Mengjiao Ren[1], Chengpeng Ji[1], Xueyan Tang[1], Haishan Tian[1], Leyong Jiang [1,*], Xiaoyu Dai[2], and Xinghua Wu[3]

[1] School of Physics and Electronics, Hunan Normal University, Changsha, 410081, China;
[2] School of Physics and Electronics, Hunan University, Changsha 410082, China;
[3] Key Laboratory for Microstructural Functional Materials of Jiangxi Province, College of Science, Jiujiang University, Jiujiang, 332005, China.

Corresponding Author: * jiangly28@hunnu.edu.cn

## Abstract

In this paper, we study the sensitivity-tunable Terahertz (THz) liquid/gas biosensor in a coupling prism-three-dimensional Dirac semimetal (3D DSM) multilayer structure. The high sensitivity of the biosensor originates from the sharp reflected peak caused by surface plasmon resonance (SPR) mode. This structure achieves the tunability of sensitivity due to that the reflectance could be modulated by the Fermi energy of 3D DSM. Besides, it is found that the sensitivity curve depends heavily on the structural parameters of 3D DSM. After parameter optimization, we obtained sensitivity over 100°/RIU for liquid biosensor. We believe this simple structure provides a reference idea for realizing high sensitivity and tunable biosensor device.

# 1. Introduction

Optical biosensor is a kind of micro-nano functional device which can transform the biological signal that is not easy to measure into the optical signal that is easy to observe and measure. Due to the interaction between the light waves and the measured bioanalyte, the slight change of the characteristics of the bioanalyte is represented by the relative obvious change of the parameters of the light signal, so as to achieve the purpose of accurately identifying and detecting the characteristics of the bioanalyte or the surrounding environment. Such sensors do not require label or modified biomolecules [1], so they are widely used in biomedicine [2,3], blood detection [4], biochemical detection [5], environmental monitoring [6], food safety [7] and other aspects. In recent years, micro-nano scale biosensors have become a research hotspot in the field of biosensors, and many optical biosensor technologies have attracted wide attention, such as fluorescent [8], colorimetric [9], optical fiber [10], evanescent wave photonics [11], etc. In addition, due to the continuous pursuit of high sensitivity and simple structure of the sensor, many optical sensor structures have been proposed and studied deeply. For example, carbon nanotubes [12], microresonators [13], surface wave imaging[14] and photonic crystal band-gap[15] and so on. Compared with traditional optical biosensor schemes, surface plasmon resonance (SPR) biosensor has many unique advantages such as high sensitivity, real-time detection, easy penetration and no loss. Therefore, SPR sensing technology has been widely used in biochemical analysis [16], food safety [17], medical diagnosis [18] and other fields, with remarkable effects, such as the detection of staphylococcal enterotoxin in milk [19], drug residue detection [20, 21], real-time disease diagnosis [22], gas detection [23], etc. However, the traditional SPR structure usually uses Otto structure or Kretschmann-Raether (KR) structure based on noble metals (such as Au [24], Ag [25] and Al [26], etc.) to excite SPR. Although relatively high sensitivity biosensing can also be realized, the existence of noble metals makes the sensors of these structures still have shortcomings such as large inherent loss, limited bandwidth and insufficient dynamic adjustability, which also brings certain limitations to the application of SPR sensors. In recent years, two-dimensional material graphene has shown great application prospects in the field of sensing due to its good photoelectric characteristics, and has become a good choice to replace traditional precious metal to excite SPR. Many graphene-based SPR sensor schemes have been proposed [27, 28]. It is generally believed that biosensors based on graphene have optimistic prospects in the field of biosensing. However, due to the limitations of the preparation of graphene-based functional devices, the implementation and promotion of graphene-based sensor schemes are also faced with the limitations of process and preparation. Therefore, the exploration of SPR sensing schemes based on new excellent materials and structures is still a challenging work.

Recently, the in-depth studies of three-dimensional Dirac semimetal (3D DSM) have opened up a new avenue for novel optical biosensors [29]. 3D DSM is a kind of 3D Dirac material, which has similar electronic and optical properties to graphene, and has the possibility of dynamic regulation, while retaining the advantages of metal-like bulk structure, with a certain thickness and relatively simple preparation, and exhibits metal-like properties under certain conditions [30]. In addition, it has beyond the ultra-high charge mobility of graphene, considerable nonlinear plasma performance, longer propagation length and strong light and material interaction. Moreover, 3D DSM is easier to prepare and process than graphene thin film, so it can better make up for the lack of graphene in photoelectric devices. All these provide a new research direction and idea for us to design SPR biosensor based on 3D DSM.

Based on this, in this paper, we theoretically propose a terahertz (THz) SPR sensor based on a coupling prism-3D DSM structure. In this structure, a THz SPR sensor with relatively high sensitivity is achieved by combining coupling prism and 3D DSM to excite SPR with 3D DSM instead of conventional noble metals or graphene. In addition, we further found that the conductivity of 3D DSM can be dynamically adjusted by Fermi energy and relaxation time, thus providing a means to control the sensitivity and figure of merit (FOM), and the thickness of the 3D DSM has a very significant effect on the sensitivity of the sensor. Through appropriate parameter optimization, we found that when the structure is applied to liquid sensing, the angle sensitivity can reach more than $100\ °/\text{RIU}$ . When the structure is used for gas sensing, the sensitivity is also relatively high. We believe that SPR tunable optical biosensors based on this structure can find possible applications in biological, chemical, environmental detection and other fields.

## 2. Theoretical Model and Method

We consider a KR configuration, which consists of coupling prism and 3D DSM layer, as shown in Fig. 1. In this structure, we use Polymethylpentene (TPX) as the coupling

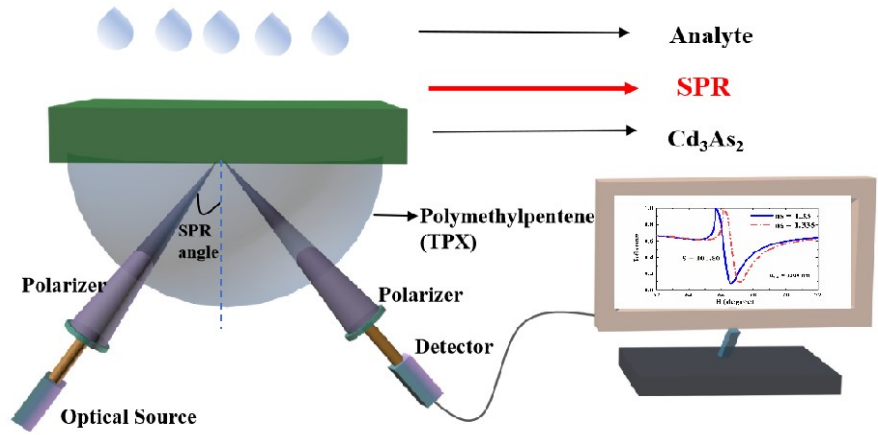

Fig. 1. Schematic representation of SPR sensor based on coupling prism and 3D DSM.

prism, with a refractive index of $n_p = 1.46$ [31], and the 3D DSM layer with candmium arsenide ($Cd_3As_2$), whose refractive index and thickness are denoted as $n_{DSM}$ and $d_{DSM}$ respectively. It is known that the conductivity of 3D DSM is electrically tunable, so by adding 3D DSM between the coupling prism and the sensing layer, the sensing performance of the whole structure can be dynamically tunable. For the 3D DSM, under the relaxation time approximation, there is a semiclassical Boltzmann transport equation, namely $\tau(\varepsilon_k) = \tau$, therefore, linear optical conductivity and refractive index of 3D DSM can be expressed as [32] :

$$\sigma_{DSM}(\omega) = \sigma_0 \frac{4}{3\pi^2} \frac{\tau}{1-i\omega\tau} \frac{(k_B T)^2}{\hbar^2 v_F} \left[ 2Li_2(-e^{-\varepsilon_F/k_B T}) + (\frac{\varepsilon_F}{k_B T})^2 + \frac{\pi^2}{3} \right] \quad (1)$$

$$n + ik = \sqrt{1 + i\sigma_{DSM}/\varepsilon_0 \omega}, \quad (2)$$

where, $\omega$ is the angular frequency of the incident light, $E_F$ is the Fermi energy, $\tau$ represents relaxation time, $\hbar$ is the reduced Planck constant, $k_B$ and $T$ are Boltzmannconstant and temperature respectively, $v_F$ is Fermi velocity, $Li_s(z)$ is the polylogarithm, and $\sigma_0 = e^2/4\hbar$. In the next calculation, the original parameter of the

3D DSM is set as $E_F = 0.1\,\text{eV}$ and $\tau = 1.0\,\text{ps}$, and the incident light frequency is 1 THz.

In order to obtain the reflectance of the whole structure, the relatively mature transfer matrix method is adopted in this scheme. We know that SPR can only be excited under TM polarization. Therefore, we only need to consider the case of TM polarization. Firstly, the transfer matrices at the junction between the coupling prism and 3D DSM and at the junction between 3D DSM and the upper sensing layer in the structure shown in Fig. 1 are respectively:

$$D_{p \to d} = \frac{1}{2}\begin{bmatrix} 1+\eta_1+\xi_1 & 1-\eta_1-\xi_1 \\ 1-\eta_1+\xi_1 & 1+\eta_1-\xi_1 \end{bmatrix}, \tag{3}$$

$$D_{d \to s} = \frac{1}{2}\begin{bmatrix} 1+\eta_2+\xi_2 & 1-\eta_2-\xi_2 \\ 1-\eta_2+\xi_2 & 1+\eta_2-\xi_2 \end{bmatrix}, \tag{4}$$

where, $\eta_1 = \varepsilon_p k_{dz}/\varepsilon_d k_{pz}$, $\xi_1 = \sigma k_{dz}/\varepsilon_0 \varepsilon_d$, $\eta_2 = \varepsilon_d k_{sz}/\varepsilon_s k_{dz}$, $\xi_2 = \sigma k_{sz}/\varepsilon_0 \varepsilon_s$. And $k_{iz}$ denotes the component of the wave vector $k_i$ in the $z$ direction, $k_i = \sqrt{\varepsilon_i}\omega/c$. $c$ is the speed of light in vacuum, $\varepsilon_0$ is the vacuum dielectric constant and $\theta$ is the incident angle. When an electromagnetic wave propagates in a uniform 3D DSM medium with thickness of $d_{DSM}$, it can be expressed as the following propagation matrix:

$$P_{DSM}(d_{DSM}) = \begin{bmatrix} e^{-ik_z d_{DSM}} & 0 \\ 0 & e^{-ik_z d_{DSM}} \end{bmatrix}. \tag{5}$$

For the structure shown in Fig. 1, the transfer matrix of the whole structure can be expressed as:

$$M = D_{p \to d} P_{DSM} D_{d \to s}, \tag{6}$$

as a result, the reflected coefficient is: $r = M_{21}/M_{11}$. Based on the above, we can

finally get the reflectance of the structure as follows:

$$R_p = |r|^2. \tag{7}$$

For a sensor, the core indicator to measure its performance includes sensitivity, half-wave full width (FWHM), the FOM and so on. Since this paper mainly studies the slight shift of the angle corresponding to the SPR reflected peak caused by the slight variation of the refractive index of the sensing layer, the angle sensitivity is used to measure the sensor, which is specifically expressed as:

$$S_\theta = \frac{\Delta \theta}{\Delta n}, \tag{8}$$

where, $\Delta \theta$ is the offset of formant angle, and $\Delta n$ represents the change of refractive index. In addition, for FOM, its calculation expression is: $FOM = S_\theta \cdot DA$, where $DA = 1/FWHM$.

## 3. Results and Discussions

In this section, we discuss the sensing characteristics of the SPR THz biosensor based on the structure shown in Fig. 1. As we know, it is a standard method for SPR biosensors to perceive small changes in the characteristics of sensing media (such as refractive index, etc.) by observing the changes of reflected peaks. In this paper, we assume that the electromagnetic wave incident from the prism at the angle of $\theta$, and plot the change of reflectance with the angle of incidence when there is 3D DSM or not, as shown in Fig. 2. It can be found from the Fig. 2 that when there is no 3D DSM, total internal reflection occurs in the partwhere $\theta$ is greater than $66°$, while after the addition of 3D DSM, a relatively obvious reflected peak can be observed nearby $\theta = 66.5°$, which is caused by SPR excitation. The confirmation of SPR can be obtained by the following dispersion relation [32]:

$$\coth\left(\sqrt{q^2 - \varepsilon_{DSM}\omega^2/c^2}\,\frac{d_{DSM}}{2}\right) = -\frac{\varepsilon_{DSM}}{\varepsilon_d}\frac{\sqrt{q^2 - \varepsilon_d\omega^2/c^2}}{\sqrt{q^2 - \varepsilon_{DSM}\omega^2/c^2}}, \quad (9)$$

where $d_{DSM}$ represents the thickness of 3D DSM. Since the dispersion curve has been reflected in other literature, it will not be described here. When the structure is used for liquid sensing, we assume that the semi-infinite background material above the 3D DSM corresponds to an aqueous solution with a refractive index of $n_s = 1.33$. Such an aqueous solution can be realized in practical experiments by constructing a flow pool above the 3D DSM layer. $Cd_3As_2$ is adopted for the 3D DSM layer, corresponding to the original thickness, Fermi energy and relaxation time, respectively $d_{DSM} = 1300$ nm, $E_F = 0.1$ eV and $\tau = 1.3$ ps. At this time, it can be clearly observed that a sharp reflected peak appears in the reflectance curve near $66.61°$, which is a typical feature of SPR excitation. When the refractive index of the sensing medium changes slightly due to the change of the environment, the reflected peak will change accordingly. We adopt the widely used refractive index variation $\Delta n = 0.005$, that is, when the refractive index changes to $n_s = 1.335$, the reflected peak moves from the initial $66.61°$ to $67.12°$, and a minor change in refractive index will cause a slight shift in the reflected peak. According to the expression (8) mentioned in the previous section, we can calculate that the sensor sensitivity can reach $102\ °/RIU$ at this time. It can be seen that the reflected peak in the reflectance curve of the structure is very sensitive to small changes in the refractive index of the sensing layer, and the SPR sensor scheme based on the structure is feasible.

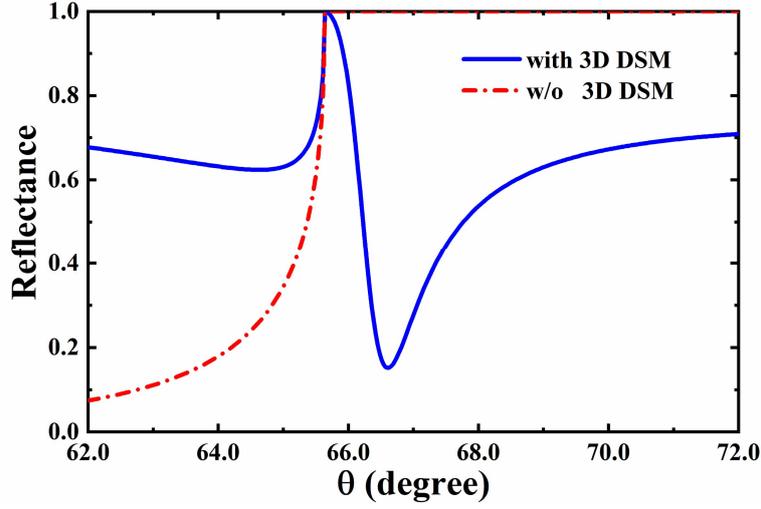

Fig. 2. Comparison plot of reflectance with incident angle with (blue line) and without 3D DSM (red line).

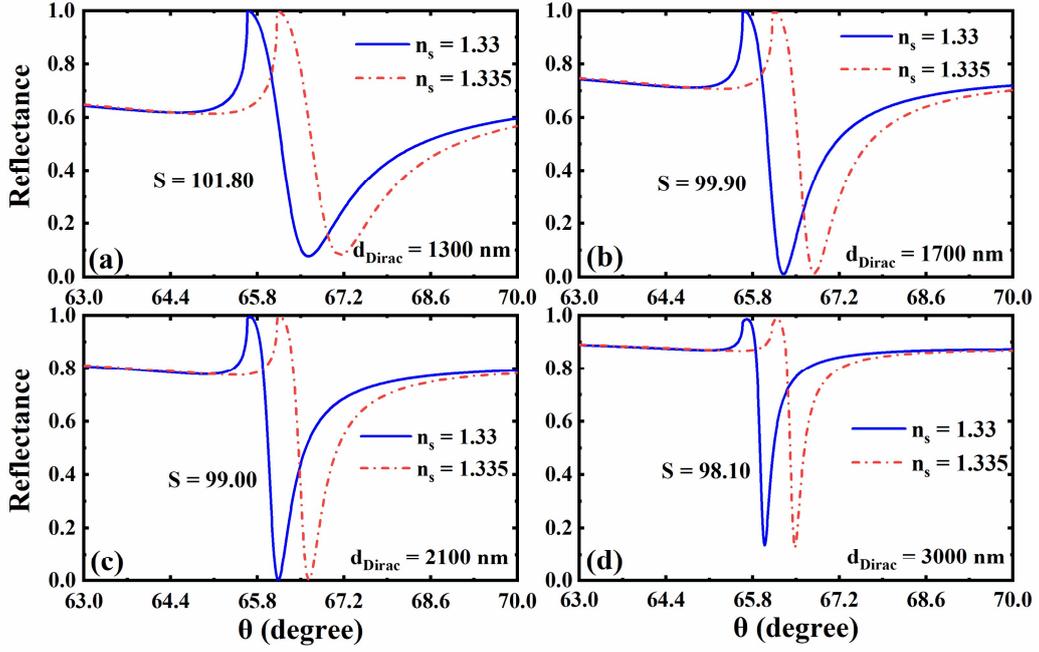

Fig. 3. The reflectance curve with the incident angle when the thickness of 3D DSM is taken as $d_{DSM} = 1300$ nm (a), $d_{DSM} = 1700$ nm (b), $d_{DSM} = 2100$ nm (c) and $d_{DSM} = 3000$ nm (d) respectively.

Combined with the previous theory, we know that the reflected peak of SPR is related not only to the structural parameters of the material, but also to the characteristics of the material itself. And the influence of the change of these parameters on the sensitivity plays a very critical role, which provides a key basis for us to design a reference scheme with higher sensitivity, so we focus on the influence of each material and structural parameter on the sensitivity of the biosensor in the following discussion. First, we consider the reflectance curve and its sensing performance under different thicknesses of 3D DSM. We have plotted the curve of the sensor reflectance as a function of the incident angle when the thickness of 3D DSM is $d_{DSM}=1300$ nm, $d_{DSM}=1700$ nm, $d_{DSM}=2100$ nm, and $d_{DSM}=3000$ nm, as shown in Fig. 3. It can be found that the reflectance curves corresponding to the thickness of 3D DSM varying within this range all have corresponding reflected peaks, that is, the excitation of SPR will not be effected, but the sharpness is slightly different, which will effect the FWHM of the reflectance curve, and then effect the FOM of the sensor. Therefore, on this basis, we plotted the changing trend curve of various related sensor performance parameter corresponding to the gradual change of 3D DSM's thickness from $d_{DSM}=1300$ nm to $d_{DSM}=3000$ nm, as shown in Fig. 4.

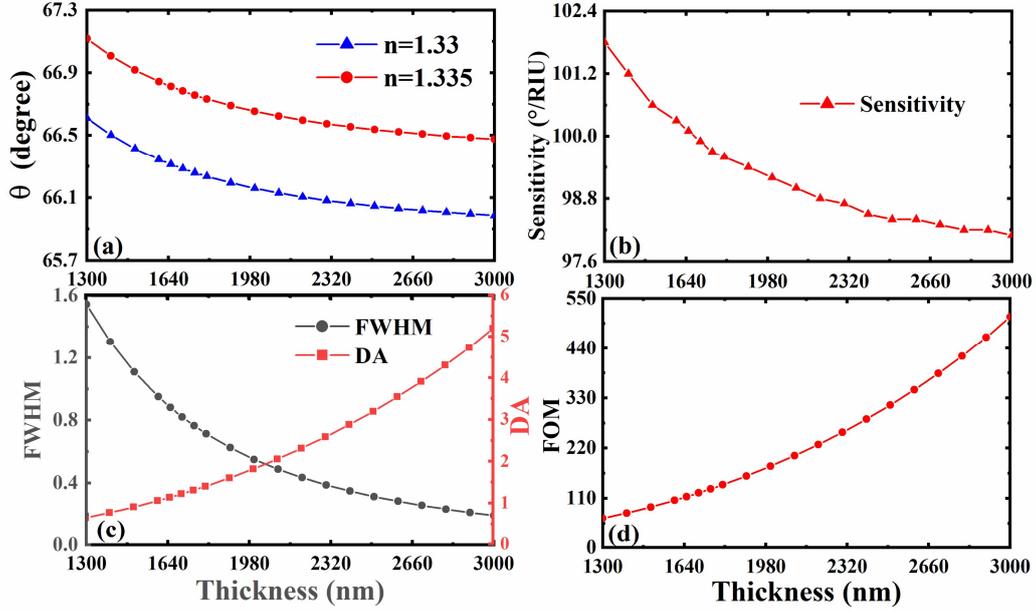

Fig. 4. The incident angle corresponding to the reflected peak (a), sensor sensitivity (b), FWHM and DA (c), FOM (d) with different thickness of 3D DSM.

It can be clearly seen from the curve results that when the thickness of 3D DSM is within the range of 1300~3000 nm, the FOM rises and the sensitivity decreases with the increase of the thickness, which has a good guiding role for our subsequent parameter selection. According to the above analysis, lower thickness of 3D DSM will make the sensor exhibit higher sensitivity. Therefore, we try to take a thinner 3D DSM, and to facilitate the calculation, we fixed its thickness to be $d_{DSM} = 1300$ nm in the subsequent design. On this basis, we will seek other regulation laws.

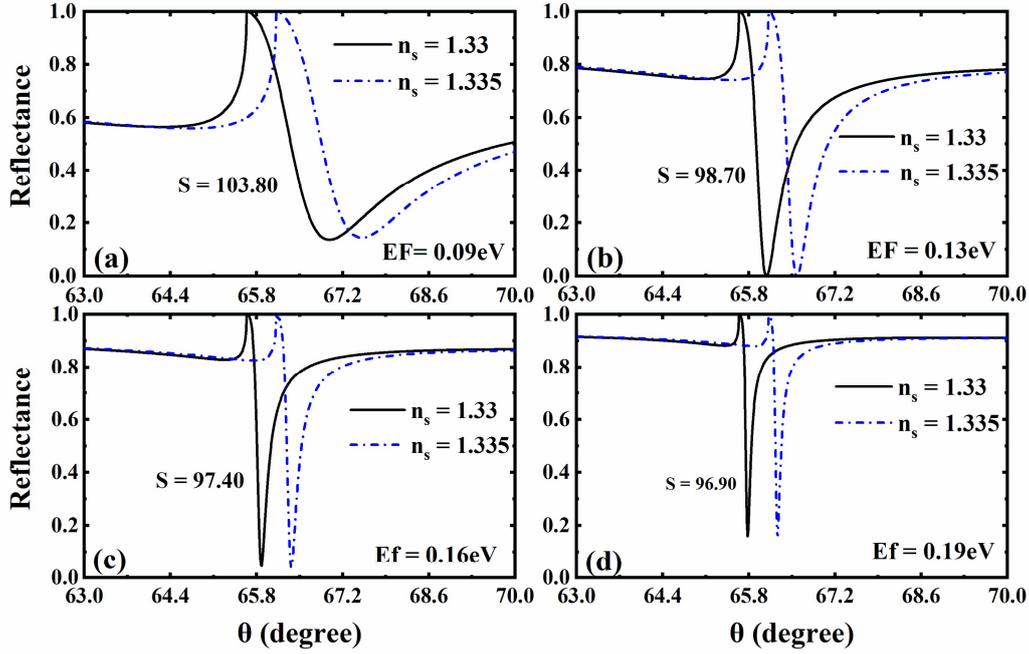

Fig.5. Curve of reflectance as a function of incident angle when the Fermi energy of 3D DSM is taken as $E_F = 0.09$ eV (a), $E_F = 0.13$ eV (b), $E_F = 0.16$ eV (c), and $E_F = 0.19$ eV (d), respectively.

Next, we further discuss the influence of material parameter properties of 3D DSM on the sensing properties of the whole structure. Material parameter properties can not only help us to find relatively high sensitivity sensing schemes, but also provide a very key reference for achieving dynamic and tunable sensitive characteristics. Then, we first pay attention to the impact of the change of 3D DSM Fermi energy on the sensing performance of the whole structure, as shown in Fig. 5.

When the Fermi energy is $E_F = 0.09$ eV, $E_F = 0.13$ eV, $E_F = 0.16$ eV, $E_F = 0.19$ eV,

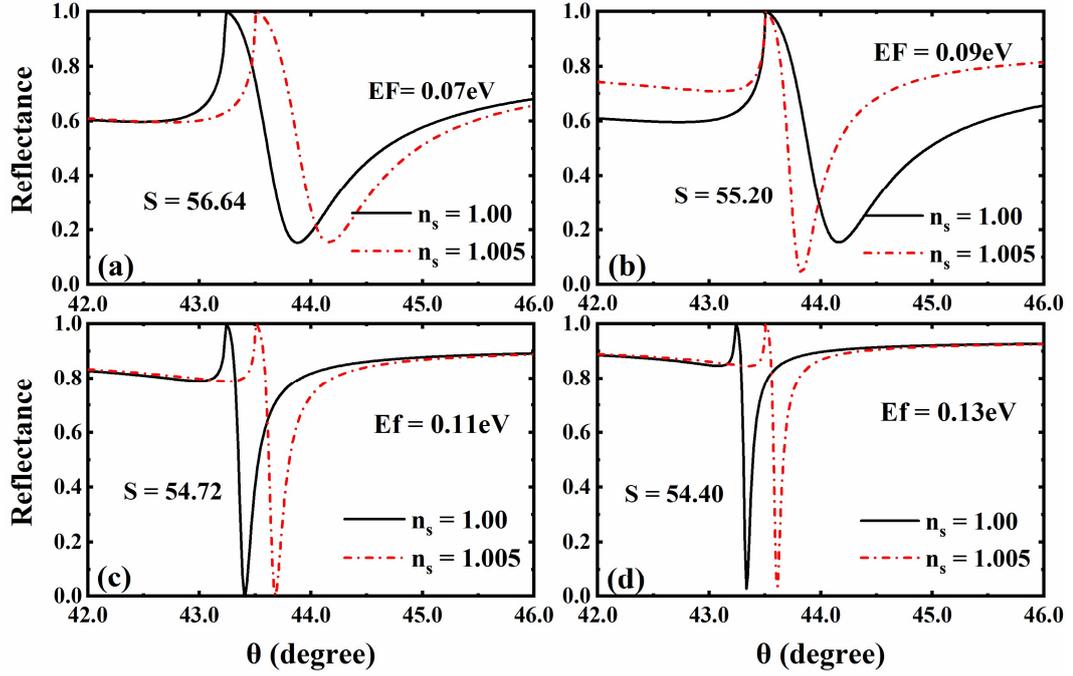

Fig.6. Curve of the reflectance with the angle of incidence at the Fermi energy of the 3D DSM $E_F = 0.07$ eV (a), $E_F = 0.09$ eV (b), $E_F = 0.11$ eV (c), $E_F = 0.13$ eV (d), respectively.

the curves of the reflectance changing with the incident angle are obtained, and the corresponding sensor sensitivity at this Fermi energy is also calculated. As can be seen from the Fig. 5, its sensitivity decreases with the increase of Fermi energy in the range of 0.09~0.19eV. However, by observing the shape of its reflected peak, the reflected peak is sharper when $E_F = 0.13$ eV, so we adjust the Fermi energy to $E_F = 0.13$ eV. In addition, we found in the calculation that the trend curve of the sensor performance changing with Fermi energy is similar to that of different 3D DSM's thicknesses (Fig. 4), and the relevant change law can be referred to in Fig. 4.

In the above discussion, we selected the parameters of the sensor for liquid, through parameter optimization, we can further explore the sensing performance when the sensing medium is gas. We set the original parameters as $d_{DSM} = 1300$ nm, $\tau = 1.3$ ps, and consider the sensitivity of the gas sensor when the refractive index of the sensing medium changes from $n_s = 1.00$ to $n_s = 1.005$. Here, we only considered the influence of Fermi energy, and plotted a comparison of reflectance curves at different Fermi energy, as shown in Fig. 6. After comparison, we found that the incident angle corresponding to the reflected peak of the gas sensor decreases compared with that of the liquid sensor. At the same time, with the increase of the Fermi energy, the reflected peak becomes sharper and sharper, while the corresponding sensor sensitivity shows a trend of gradual decrease. Moreover, after certain parameter optimization, we can obtain a gas sensor with better performances.

## 4. Conclusions

In summary, we propose a THz sensor scheme based on SPR. This scheme excite the SPR by a coupling prism and 3D DSM, generating sharp reflected peaks. Meanwhile the introduction of 3D DSM significantly improves the sensor sensitivity and it also provides a means to dynamically adjust its sensing characteristics. The calculation results show that the sensing performance of the SPR biosensor is not only related to the structural parameters, but also closely related to the material parameters of 3D DSM. When the sensor is applied to liquid sensing, through structure and parameter optimization, we can obtain the sensitivity of over $100 \ °/RIU$, and at the same time, when the structure is used for gas sensing, good sensing performance is also obtained. Compared with the traditional SPR biosensor, this scheme has a more simple structure, a lower production process requirements, and a relatively high sensitivity. We believe this scheme is expected to show potential applications in the field of micro-nano structure-based biosensing.

## Acknowledgments

This work was supported by the Scientific Research Fund of Hunan Provincial Education Department (Grant No. 21B0048), Natural Science Foundation of Hunan Province (Grant Nos.2022JJ30394), the Changsha Natural Science Foundation (Grant Nos. kq2202236, kq2202246), and the Science and Technology Project of Jiangxi Provincial Education Department (Grant No. GJJ190911).

## References

1. Fan, X.; White, I. M.; Shopova, S. I.; Zhu, H.; Suter, J.D.; Sun, Y. Sensitive optical biosensors for unlabeled targets : A review. *Anal. Chim. Acta.* **2008,** 620, 8-26.
2. Son, M. H.; Park, S. W.; Sagong, H. Y.; Jung, Y. K. Recent Advances in Electrochemical and Optical Biosensors for Cancer Biomarker Detection. *Biochip. J.* **2022**, 17, 44-67.
3. Servarayan, K. L.; Sundaram, E.; Manna, A.; Sivasamy, V. V. Label free optical biosensor for insulin using naturally existing chromene mimic synthesized receptors: A greener approach. *Anal. Chim. Acta.* **2023,** 1239, 340692.
4. Hamouleh-Alipour, A.; Forouzeshfard, M.; Baghbani, R.; Vafapour, Z. Blood Hemoglobin Concentration Sensing by Optical Nano Biosensor-Based Plasmonic Metasurface: A Feasibility Study. *IEEE. Trans. Nanotechnol.***2022,** 21, 620-628.
5. Sanz, V.; Marcos, S. D.; Galbán, J. A reagentless optical biosensor based on the intrinsic absorption properties of peroxidase. *Biosens. Bioelectron.***2007,** 22, 956-964.
6. Sivakumar, R.; Lee, N. Y. Recent advances in airborne pathogen detection using optical and electrochemical biosensors. *Anal.Chim.Acta.* **2022,** 1234, 340297.
7. Pebdeni, A. B.; Roshani, A.; Mirsadoughi, E.; Behzadifar, S.; Hosseini, M. Recent advances in optical biosensors for specific detection of E. coli bacteria in food and water. *Food Control.* **2022,** 135, 108822.

8. Yang, Y.; Lei, X.; Liu, B.; Liu, H.; Chen, J.; Fang, G.; Liu, J.; Wang, S. A ratiometric fluorescent sensor based on metalloenzyme mimics for detecting organophosphorus pesticides. *Sens. Actuators. B. Chem.* **2023,** 377, 133031.

9. Zhu, J.; Yang, B.; Hao, H.; Peng, L.; Lou, S. Gold nanoparticles-based colorimetric assay of pesticides: A critical study on aptamer's role and another alternative sensor array strategy. *Sens. Actuators. B. Chem.* **2023,** 381, 133439.

10. Li, X.; Gong, P.; Zhang, Y.; Zhou, X. Label-Free Micro Probe Optical Fiber Biosensor for Selective and Highly Sensitive Glucose Detection. *IEEE. Trans. Instrum. Meas.* **2022,** 7, 1-8.

11. Huertas, C. S.; Calvo-Lozano, O.; Mitchell, A.; Lechuga, L. M. Advanced Evanescent-Wave Optical Biosensors for the Detection of Nucleic Acids: An Analytic Perspective. *Front. Chem.* **2019,** 7, 724.

12. Shumeiko, V.; Malach, E.; Helman, Y.; Paltiel, Y.; Bisker, G.; Hayouka, Z.; Shoseyov, O. A nanoscale optical biosensor based on peptide encapsulated SWCNTs for detection of acetic acid in the gaseous phase. *Sens. Actuators. B. Chem.* **2021**, 327, 128832.

13. Malmir, K.; Habibiyan, H.; Ghafoorifard, H. Ultrasensitive optical biosensors based on microresonators with bent waveguides. *Optik.* **2020,** 216, 164906.

14. Konopsky, V. N.; Alieva, E. V. Imaging biosensor based on planar optical waveguide. *Opt. Laser. Technol.* **2019,** 115, 171-175.

15. Khani, S.; Hayati, M. Optical biosensors using plasmonic and photonic crystal band-gap structures for the detection of basal cell cancer. *Sci. Rep.* **2022,** 12, 1-19.

16. Puttharugsa, C.; Wangkam, T.; Houngkamhang, N.; Yodmongkol, S.; Gajanandana, O.; Himananto, O.; Sutapun, B.; Amarit, R.; Somboonkaew, A.; Srikhirin, T. A polymer surface for antibody detection by using surface plasmon resonance via immobilized antigen. *Curr. Appl. Phys.* **2013,** 13, 1008-1013.

17. BİBEROĞLU, Ö. The Determination of Listeria monocytogenes in Foods with Optical Biosensors. *Van. Vet. J.* **2020,** 31, 50-55.

18. Ermini, M. L.; Mariani, S.; Scarano, S.; Minunni, M. Bioanalytical approaches for the detection of single nucleotide polymorphisms by Surface Plasmon Resonance biosensors. *Biosens. Bioelectron.* **2014,** 61, 28-37.

19. Homola, J.; Dostálek, J.; Chen, S.; Rasooly, A.; Jiang, S.; Yee, S. S. Spectral surface plasmon resonance biosensor for detection of staphylococcal enterotoxin B in milk. *Int. J. Food. Microbiol.* **2002,** 75, 61-69.

20. Situ, C.; Crooks, S. R.; Baxter, A. G.; Ferguson, J.; Elliott, C. T. On-line detection of sulfamethazine and sulfadiazine in porcine bile using a multi-channel high-throughput SPR biosensor. *Anal. Chim. Acta.* **2002,** 473, 143-149.

21. Caldow, M.; Stead, S. L.; Day, J.; Sharman, M.; Situ, C.; Chris, E. Development and validation of an optical SPR biosensor assay for tylosin residues in honey. *J. Agric. Food. Chem.* **2005,** 53, 7367-70.

22. Riedel, T.; Rodriguez-Emmenegger, C.; Pereira, A. S.; Bědajánková, A.; Jinoch, P. Boltovets, P. M.; Brynda, E. Diagnosis of Epstein‑Barr virus infection in clinical serum samples by an SPR biosensor assay. *Biosens. Bioelectron.* **2014,** 55, 278-284.

23. Yuxin, L.; Hailiang, C.; Qiang, C.; Hongwei, L.; Zhigang, G. Surface plasmon resonance induced methane gas sensor in hollow core anti-resonant fiber. *Opt. Fiber. Technol.* **2023,** 78, 103293.

24. Lokman, N. F.; Bakar, A.; Ashrif, A.; Suja, F.; Abdullah, H.; Rahman, W. B. W. A.; Huang, N.; Yaacob, M. H. Highly sensitive SPR response of Au/chitosan/graphene oxide nanostructured thin films toward Pb (II) ions. *Sens. Actuators. B. Chem.* **2014,** 195, 459-466.

25. Leong, K. H.; Gan, B. L.; Ibrahim, S.; Saravanan, P. Synthesis of surface plasmon resonance (SPR) triggered Ag/TiO2 photocatalyst for degradation of endocrine disturbing compounds. *Appl. Surf. Sci.* **2014,** 319, 128-135.

26. Hu, M.; Li, M.; Li, M. Y.; Wen, X.; Deng, S.; Liu, S.; Lu, H. Sensitivity Enhancement of 2D Material-Based Surface Plasmon Resonance Sensor with an Al-Ni Bimetallic Structure. *Sensors.* **2023,** 23, 1714.

27. Fares, H.; Almokhtar, M.; Almarashi, J. Q. M.; Rashad, M.; Moustafa, S. Tunable narrow-linewidth surface plasmon resonances of graphene-wrapped dielectric nanoparticles in the visible and near-infrared. *Physica. E. Low. Dimens. Syst. Nanostruct.* **2022,** 142, 115300.

28. Chen, S.; Chu, S.; Song, Y.; Wu, H.; Liu, Y.; Peng, W. Near-infrared surface plasmon resonance sensor with a graphene-gold surface architecture for ultra-sensitive biodetection. *Anal. Chim. Acta.* **2022,** 1205, 339692.

29. Liu, Z. K.; Jiang, J.; Zhou, B.; Wang, Z. J.; Zhang, Y .; Weng, H. M.; Prabhakaran, D.; Mo, S. K.; Peng, H.; Dudin, P. A stable three-dimensional topological Dirac semimetal $Cd_3As_2$. *Nat. Mater.* **2014,** 13, 677-681.

30. Kotov, O. V.; Lozovik, Y. E. Dielectric response and novel electromagnetic modes in three-dimensional Dirac semimetal films. *Phys.Rev. B.***2016,** 93, 235417.

31. Leyong, J.; Jiao, T.; Qingkai, W.; Yuexiang W.; Zhiwei Z.; Yuanjiang X.; Xiaoyu, D. Manipulating optical Tamm state in the terahertz frequency range with graphene. *Chin. Opt. Lett.***2019,** 17, 20008.

32. Ooi, K.; Ang, Y.S.; Zhai, Q.; Tan, D.; Ang, L.K.; Ong, C.K. Nonlinear plasmonics of three-dimensional dirac semimetals. *APL. Photonics.***2019,** 4, 034402.